\documentclass[a4paper]{article}

\usepackage{INTERSPEECH2022}
\usepackage{mathrsfs}
\usepackage{amssymb}
\usepackage{kotex}
\usepackage{xcolor}

\usepackage{amsmath}
\DeclareMathOperator*{\argmax}{arg\,max}

\DeclareMathOperator*{\maxpool}{Maxpool}

\newcommand{\bg}[1]{\textcolor{black} {#1}} % BG
\title{Dummy Prototypical Networks for Few-Shot Open-Set Keyword Spotting}
\name{Byeonggeun Kim$^1$, Seunghan Yang$^1$, Inseop Chung$^{1,2,*}$, \& Simyung Chang$^1$}
\address{
$^1$Qualcomm AI Research${}^{\dagger}$, Qualcomm Korea YH, Seoul, Republic of Korea  \thanks{  ${}^{\dagger}$ Qualcomm AI Research is an initiative of Qualcomm Technologies, Inc.${}^{*}$Author completed the research in part during an internship at Qualcomm Technologies, Inc.}\\
$^2$Seoul National University, Seoul, Republic of Korea
}
\email{\{kbungkun, seunghan, ichung, simychan\}@qti.qualcomm.com}

\begin{document}

\maketitle
\begin{abstract}
Keyword spotting is the task of detecting a keyword in streaming audio. Conventional keyword spotting targets predefined keywords classification, but there is growing attention in few-shot (query-by-example) keyword spotting, e.g., $N$-way classification given $M$-shot support samples. Moreover, in real-world scenarios, there can be utterances from unexpected categories (open-set) which need to be rejected rather than classified as one of the $N$ classes. Combining the two needs, we tackle few-shot open-set keyword spotting with a new benchmark setting, named \textit{split}GSC. We propose episode-known dummy prototypes based on metric learning to detect an open-set better and introduce a simple and powerful approach, Dummy Prototypical Networks (D-ProtoNets). Our D-ProtoNets shows clear margins compared to recent few-shot open-set recognition (FSOSR) approaches in the suggested \textit{split}GSC. We also verify our method on a standard benchmark, \textit{mini}ImageNet, and D-ProtoNets shows the state-of-the-art open-set detection rate in FSOSR.
\end{abstract}

\section{Introduction}

Keyword spotting (KWS) detects keywords like ``Hey, Google'' and ``Hey, Siri'' in streaming audio. KWS systems usually target edge devices such as mobile phones and smart speakers, and previous studies have concentrated on better network designs in terms of detection rate~\cite{res15, orthgonal_attn_kws} and computational cost~\cite{ds-resnet,bcresnet} while targeting multi-keyword classifications with Google speech commands dataset (GSC) version 1 and 2~\cite{GSCdataset}.

Recently, there has been growing attention in query-by-example (few-shot) keyword spotting systems~\cite{QbyE_KWS_Qualcomm_dataset, fewshotKWS_interspeech20, FSL_KWS_protonet20, QbyEKWS_att_icassp21}. Few-shot learning (FSL) absorbs knowledge from a training dataset and leverages the knowledge to adapt to evaluation tasks of unseen categories using only a few labeled (support) samples. However, on top of FSL, real-world scenarios naturally meet utterances of unexpected categories without support examples, and neural networks tend to be over-confident~\cite{foolNN} and can misjudge those unexpected samples by one of the FSL classes. Thus, there is a need to detect those unseen \textit{open-set} classes (Open-Set Recognition (OSR)~\cite{OSR}). This work introduces OSR to few-shot keyword spotting that is a more challenging setting, few-shot open-set recognition (FSOSR)~\cite{PEELER} for KWS.

\begin{figure}[t]
  \centering
  \includegraphics[width=0.95\linewidth]{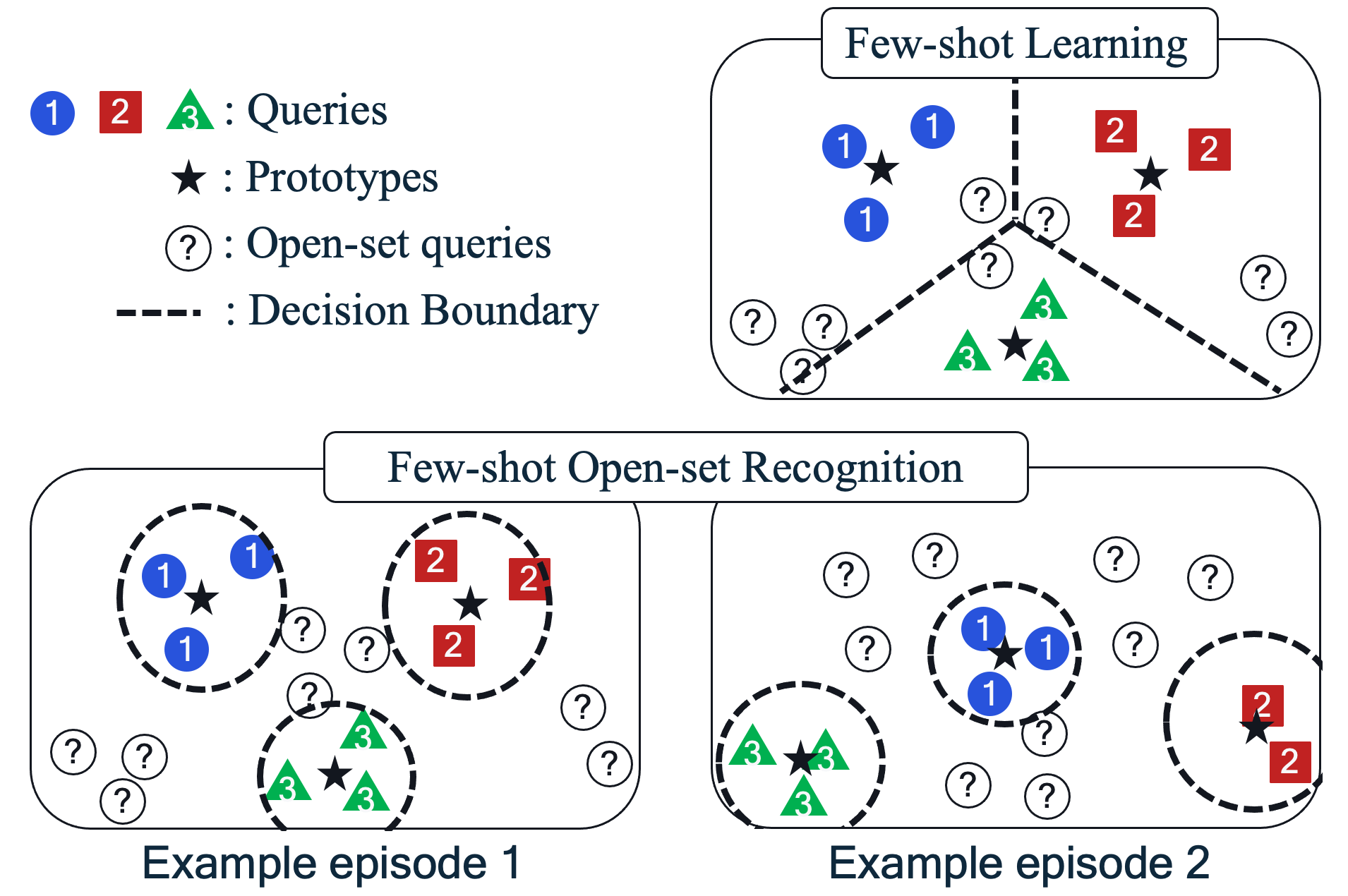}
  \vskip -0.15in
  \caption{\textbf{Example episodes:} In few-shot learning, decision boundaries classify classes given few support samples. Few-shot open-set recognition is required to distinguish samples of unknown categories (open-set) from those of known classes. Open-set can vary over episodes in FSOSR.
  }
  \vskip -0.3in
  \label{fig:fsosr}
\end{figure}

Figure~\ref{fig:fsosr} shows example episodes of FSL and FSOSR while using an FSL method, Prototypical Networks (ProtoNets)~\cite{protonet}. In an $N$-way $M$-shot episode, the goal of FSL is correctly classifying $N$ classes that are unseen during training but \textit{known} using $M$ support samples for each. FSL does not consider open-set classes out of the $N$ classes. On the other hand, FSOSR needs to distinguish an \textit{unknown} open-set from the known classes while still performing FSL. FSOSR is more challenging than conventional OSR because an open-set changes over episodes based on the choice of $N$ classes (Figure~\ref{fig:fsosr} bottom). Thus, a desirable FSOSR method needs to adapt to the varying open-set. We predict episode-specific (\textit{episode-known}) dummies based on support examples in each episode and classify an open-set as the dummies. Using the episode-known dummies, we propose Dummy Prototypical Networks (D-ProtoNets).

For few-shot open-set keyword spotting (FSOS-KWS), we introduce a benchmark setting named \textit{split}GSC, a subset of GSC ver2. Our D-ProtoNets achieves state-of-the-art (SOTA) performance in \textit{split}GSC. We also verify D-ProtoNets on \textit{mini}ImageNet~\cite{matchingnet_conv4_64_miniimagenet}, a widely used FSL benchmark, and D-ProtoNets is better in detecting open-set than other baselines.
% \begin{itemize}
%     \item We suggest a benchmark-setting for few-shot open-set keyword spotting (FSOS-KWS), named \textit{miniGSC}.
%     \item We introduce episode-known dummy classes to Prototypical Networks~\cite{protonet}, named Dummy Prototypical Networks (D-ProtoNets), and it achieves SOTA in \textit{split}GSC and better open-set detection in \textit{mini}ImageNet.
% \end{itemize}

\begin{figure*}[t]
  \centering
  \includegraphics[width=0.9\linewidth]{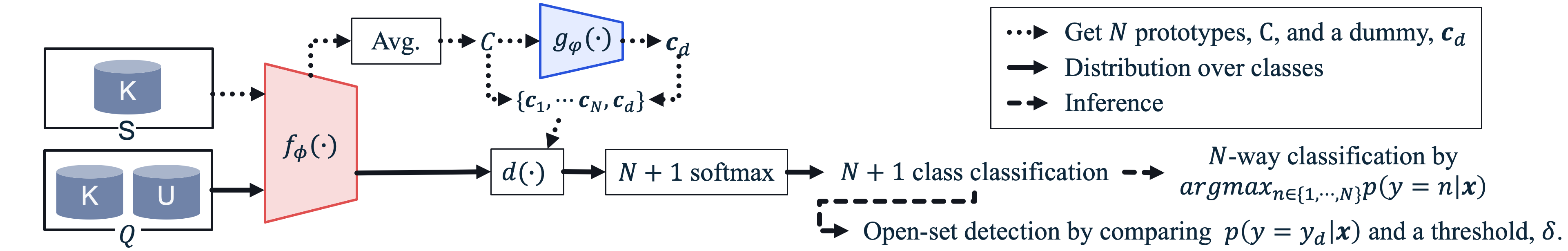}
  \vskip -0.13in
  \caption{\textbf{Dummy Prototypical Network.} $f_\phi(\cdot)$ and $g_\varphi(\cdot)$ are the encoder and the dummy generator, respectively. `K' and `U' stand for a set of known classes having support samples and unknown open-set, respectively, and $d(\cdot)$ is a distance metric.
  }
  \vskip -0.25in
  \label{fig:overall_system}
\end{figure*}

\section{Related Works}

The literature on few-shot learning and open-set recognition are vast, thus we focus on the most relevant works. 

\noindent \textbf{Few-Shot Learning.} Few-shot learning has three popular branches, adaptation, hallucination, and metric learning methods. The adaptation methods~\cite{MAML} make a model easy to fine-tune in the low-shot regime, and the hallucination methods~\cite{hallucination} augment training examples for data starved classes. Our approach aligns with the last one, metric-based learning~\cite{protonet,matchingnet_conv4_64_miniimagenet}, which learns a metric space in which distance metrics can classify samples. Especially our method is designed on top of Prototypical Networks (ProtoNets)~\cite{protonet}. Recently, FEAT~\cite{FEAT} shows that it is helpful to make support samples task-specific using a set-to-set function, Transformer~\cite{transformer}, in FSL. Also, some approaches have addressed few-shot KWS~\cite{fewshotKWS_interspeech20, FSL_KWS_protonet20, QbyEKWS_att_icassp21}, but to the best of our knowledge, this is the first work introducing few-shot open-set keyword spotting (FSOS-KWS).

\noindent \textbf{Open-Set Recognition.} \cite{OSR} introduced OSR to deep learning, and discriminative or generative approaches have been proposed by \cite{CROSR, GOpenMAx, gen_openset2}. Recently, \cite{PROSER} suggests placeholders for both data and classifiers using manifold mixup~\cite{manifold_mixup} and learnable dummy classifiers, respectively. Their dummy classifiers are learnable, fixed, and thus suitable for OSR with fixed closed-set. On the other hand, our D-ProtoNets suggests episode-known, varying dummies for FSOSR.

\noindent \textbf{Few-shot open-set recognition.} There is growing attention to the FSOSR due to its importance, and the previous studies~\cite{PEELER, snatcher_cvpr21} concentrate on better OSR while preserving FSL performance. PEELER~\cite{PEELER} suggests entropy maximization loss for an open-set and Gaussian Embedding for flexible decision boundaries. Based on a set-to-set transformation method~\cite{FEAT}, \cite{snatcher_cvpr21} introduces SnaTCHer, the SOTA FSOSR approach that compares the distance between transformed and modified prototypes and detects open-set using transformation consistency. Those works try to detect abnormality of open-set while our method directly learns a dummy class to detect open-set.

\section{Method}

\subsection{Preliminaries}

\noindent \textbf{Notations.} An FSOSR setting consists of \textit{seen} training data $\mathcal{D}_\text{train}$ and \textit{unseen} evaluation data $\mathcal{D}_\text{eval}$ that do not overlap classes. $\mathcal{D}_\text{train}$ is composed of labeled samples, $\{(\bm{x}_i, y_i)\}_{i=1}^{|\mathcal{D}_\text{train}|}$, where $\bm{x_i}$ is an input feature, and $y_i$ is its corresponding label. During training, a model learns from $N$-way $M$-shot pseudo-FSOSR episodes, each of which has $N$ \textit{known} classes offering $M$ support examples for each class and \textit{pseudo-unknown} (pseudo-open-set) classes without any support examples. We denote a support set and a query set as $S$ and $Q$, respectively, in an episode. $S$ contains samples from $N$ classes, and $Q$ contains queries from both $N$ known and $N_U$ unknown classes where $|S|=NM$, $|Q| = (N + N_U) M_Q$, and $M_Q$ is the number of queries for each class. At inference time, all classes of $\mathcal{D}_\text{eval}$ are unseen, and again an episode consists of $N$ known classes with support samples and $N_U$ unknown open-set classes.

\noindent \textbf{Prototypical Networks.} Our work is in line with metric-learning-based approaches and is based on the representative work, Prototypical Networks (ProtoNets)~\cite{protonet}. In an N-way M-shot episode, ProtoNets gets $N$ prototypes, $\{\bm{c}_n\}_{n=1}^N$, using the average of the support samples of each class, $n$
, where
\begin{equation}
\label{eq:prototype}
\bm{c}_n = \frac{1}{M}\sum_{(\bm{x}_i, y_i)\in S_n}{f_\phi(\bm{x}_i)},
\end{equation}
$S_n$ is a subset of $S$ whose labels are $n$, $|S_n|=M$, and $f_\phi$ is an encoder with $\phi$ parameters. Based on the prototypes, ProtoNets gets distribution over $N$ classes,
\begin{equation}
\label{eq:protonet}
p_\phi(y=n|\bm{x}) = \frac{\exp(-d(f_\phi(\bm{x}), \bm{c}_n))}{\sum_{n'=1}^{N}{\exp(-d(f_\phi(\bm{x}) ,\bm{c}_{n'})})},
\end{equation}
where $d(\cdot)$ is a distance metric, e.g., Euclidean distance, $d(z, z') = ||z-z'||^2$. Using Eq.~\ref{eq:protonet}, ProtoNets minimizes negative log-probability, $- \log  p_\phi(y=n|\bm{x})$, of the true class $n$.

\subsection{Few-shot Open-set Keyword Spotting}
\label{sec:FSOS-KWS setting}
Here, we introduce a benchmark-setting of FSOS-KWS using Google speech commands dataset (GSC) ver2~\cite{GSCdataset}. There are 35 keywords in total, and conventional KWS does 12 class classifications following the settings of \cite{GSCdataset}. The 12 classes consist of 10 keywords ( ``Yes," ``No,” ``Up,” ``Down,” ``Left,” ``Right,” ``On,” ``Off,” ``Stop,” and ``go'') and two additional classes ``Unknown words'' which refers to the remaining 25 keywords, and ``Silence'' (background noise only). We split the dataset by class label like \textit{mini}ImageNet~\cite{matchingnet_conv4_64_miniimagenet}. Our split has 15, 10, and 10 keywords for train, validation, and test set, respectively, as follows:
\begin{itemize}
    \item Train keywords: ``Happy,'' ``House,'' ``Bird,'' ``Bed,'' ``Backward,'' ``Sheila,'' ``Marvin,'' ``Wow,'' ``Tree,'' ``Follow,'' ``Dog,'' ``Visual,'' ``Forward,'' ``Learn,'' and ``Cat''.
    \item Validation keywords (numbers): ``Zero,'' ``One,'' ``Two,'' ``Three,'' ``Four,'' ``Five,'' ``Six,'' ``Seven,'' ``Eight,'' and ``Nine''.
    \item Test keywords (10 keyword classes used in conventional 12 class KWS): ``Yes," ``No,” ``Up,” ``Down,” ``Left,” ``Right,” ``On,” ``Off,” ``Stop,” and ``go''.
\end{itemize}
These fixed keyword splits can prevent possible performance variance from split changes over trials. On top of the split, we add the particular class ``Silence'' which can only be included in an open-set as a background noise class. For example, in a 5-way 5-shot episode, we randomly choose five known classes without ``Silence'' and then choose the same number of open-set classes from the remaining classes, including ``Silence''. We name this specific setting as \textit{split Google speech commands dataset} (\textit{split}GSC). More details are available in Section~\ref{sec:experimentalsetting}.

\subsection{Dummy Prototypical Network}

We suggest \textit{episode-known} dummy prototypes and introduce a simple but powerful system named \textit{Dummy Prototypical Networks} (D-ProtoNets) to handle varying open-sets over episodes.

\noindent \textbf{Episode-known Dummy Prototype.} Let us consider $N$ prototypes, $\{\bm{c}_n\}_{n=1}^N$, in an episode, where $\bm{c} \in R^{1\times D}$ with the output dimension of $f_\phi$, $D$. The prototypes are permutation invariant each other for ProtoNets. Thus, we suggest a dummy generator, $g_\varphi$, with parameters $\varphi$ based on DeepSets~\cite{deepsets}, which is inherently permutation invariant. In detail, $g_\varphi$ generates a dummy $\bm{c}_d$ using given $N$ prototypes $C=[\bm{c}_1;\bm{c}_2;\cdots; \bm{c}_N]\in R^{N\times D}$ as an input, hence episode-known:
\begin{equation}
\label{eq:dummy_generator}
\bm{c}_d = g_\varphi(C) = \maxpool(g_1(C))W_g,
\end{equation}
where $g_1$ consists of fully-connected (FC) layers with nonlinearity, and $g_1(C)\in R^{N\times H}$ with a hidden dimension $H$. $\maxpool$ operates over $N$ and outputs a feature in $R^{1\times H}$. $W_g$ is a learnable $H \times D$ matrix, and $\bm{c}_d\in R^{1\times D}$. Finally, we get an augmented prototype set $\{\bm{c}_1, \cdots, \bm{c}_N, \bm{c}_d\}$. We set the labels of open-set queries to the N+1-th label $y_d$ which corresponds to the dummy, $\bm{c}_d$. Then, the Eq.~\ref{eq:protonet} becomes 
\begin{equation}
\label{eq:dummy_protonet}
p_{\theta}(y=n|\bm{x}) = \frac{\exp(-d(f_\phi(\bm{x}), \bm{c}_n)/\tau_n)}{\sum_{n'=1}^{N+1}{\exp(-d(f_\phi(\bm{x}) ,\bm{c}_{n'})/\tau_n)}},
\end{equation}
where $\theta$ consists of the encoder $\phi$ and the dummy generator $\varphi$, and $\tau_n$ is a softmax temperature which is usually same over classes. Here we use a larger $\tau_{N+1}$ compared to other $\tau_{n \neq N+1}$ to let the dummy easily reduce its loss, i.e., $\tau_{N+1}=\gamma\cdot \tau_{n \neq N+1}$, where $\gamma > 1$. The N+1 classification is learned by cross entropy loss as below:
\begin{align}
\label{eq:celoss}
\mathcal{L}^K_{CE}=\sum_{(\bm{x}_i, y_i)\in Q_K}{-\log p_{\theta}(y=y_i|\bm{x}_i)}, \nonumber\\
\mathcal{L}^U_{CE}=\sum_{(\bm{x}_i, y_d)\in Q_U}{-\log p_{\theta}(y=y_d|\bm{x}_i)},
\end{align}
where $Q_K$ and $Q_U$ are known and unknown queries, respectively. We balance the two losses by $\lambda$, and the total loss is $\mathcal{L}_{CE}=\mathcal{L}^K_{CE} + \lambda \cdot \mathcal{L}^U_{CE}$.

During a test, we classify $N$ known classes by $\hat{y_i} = \argmax_{n\in \{1,\cdots, N\}} p_\theta(y_i=n|\bm{x}_i)$. We verify whether a $x_i$ is an open-set by comparing $p_\theta(y_i=y_d|\bm{x}_i)$ to a threshold $\delta$. Overall system is described in Figure~\ref{fig:overall_system}.

\noindent \textbf{Multiple Dummies.} We expand D-ProtoNets with multi $L$ dummies by changing $W_g$ to a $H\times (L\cdot D)$ matrix. The model can naively choose most probable dummy for an input $\bm{x}_i$ by $\argmax_l(-d(\bm{x}_i, \bm{c}_l))$. Instead, we use Gumbel softmax~\cite{gumbel} to replace the non-differentiable sample, $\argmax$, with a differentiable sample. The probability of choosing dummy $l$ is
\begin{equation}
\label{eq:gumbel}
p(y^L=l|\bm{x}) = \frac{\exp((-d(f_\phi(\bm{x}), \bm{c}_l)+\epsilon_l)/\tau)}{\sum_{l'=1}^{L}{\exp((-d(f_\phi(\bm{x}) ,\bm{c}_{l'})+\epsilon_{l'})/\tau})},
\end{equation}
where $\epsilon_1, \cdots, \epsilon_L$ are i.i.d samples drawn from the standard Gumbel distribution of $\mu=0$ and $\beta=1$ following \cite{gumbel}, and $y^L$ is a temporal dummy label among L dummies. During training, we get $\bm{c}_d=\sum_{l}{p(y^L=l|\bm{x})\cdot \bm{c}_l}$, and we choose $y^L$ by $\argmax_lp(y^L=l|\bm{x})$ at inference time. 
% On top of that, we use a diversity loss,
% \begin{equation}
% \label{eq:diverse_loss}
% \mathcal{L}_D = \frac{1}{|Q|}\sum_{x_i\in Q}{\sum_{l=1}^{L}{p(y^L_i=l|x_i) \log p(y^L_i=l|x_i)}},
% \end{equation}
% to encourage various dummy selection. The total loss is $\mathcal{L} = \mathcal{L}^K_{CE}+\lambda\cdot \mathcal{L}^U_{CE} + \alpha\cdot \mathcal{L}_{D}$, where $\alpha$ is a weighting hyperparameter.

\section{Experiments}

\subsection{Experimental Settings}
\label{sec:experimentalsetting}
\noindent \textbf{Dataset.} We use Google speech commands (GSC) dataset ver2~\cite{GSCdataset}, containing 105,829 utterances from 2,618 speakers. The dataset is first split to train, validate, and test sets with 84,843, 9,981, and 11,005 utterances, respectively, using the official split (using a hash function on the name of each utterance file)~\cite{GSCdataset}. Then, we chose the samples based on our \textit{split}GSC split and got 22,916, 3,643, and 4,074 samples for train, validation, and test, respectively. Then following the settings of \cite{GSCdataset}, we add ``Silence'' samples to each split by the average number of utterances per class of each split, and finally, \textit{split}GSC has 24,444, 4,007, and 4,482 utterances for train, validation and test, respectively. Especially, we use the official test set that \cite{GSCdataset} offers and apply our split to it. During the training, we use minimal data augmentation, commonly used in GSC tasks~\cite{res15, bcresnet, GSCdataset}: Adding official background noise offered by GSC with the probability of 0.8.

\noindent \textbf{Backbones.} We experiment with two widely used backbones in previous FSL benchmarks~\cite{protonet, FEAT}, Conv4-64~\cite{matchingnet_conv4_64_miniimagenet} and ResNet-12~\cite{resnet12}, and one backbone designed for KWS, BCResNet-8~\cite{bcresnet}. Each corresponds to the encoder, $f_\phi$, and their output dimensions are 768, 512, and 256 for Conv4-64, ResNet12, and BCResNet-8, respectively. Conv4-64 does not have global average pooling at the end, and thus we get 768 dimensions, larger than its number of channels, 64.

\noindent \textbf{Implementation Details.} 
Each utterance in GSC is 1 sec long, and the sampling rate
is 16 kHz. We use input features of 40-dimensional log Mel-spectrograms with frameshift and window length of 10 and 30 ms, respectively, following \cite{bcresnet}. We train a model for 100 epochs with Adam optimizer~\cite{adam} of initial learning rate of 0.001. The learning rate is step decayed by multiplying 0.5 at every 20 epochs. Each epoch consists of 100 episodes, and an episode has 5 known (5-way) and 5 open-set classes. We use 5 support examples (5-shot), and all the classes have 5 and 15 queries for each during training and test, respectively. We use early-stop by few-shot validation accuracy and evaluate a trained model with 1,000 episodes. 
% We report 5-way 1-shot results by interpolating the number of shots during a test. 

We design the dummy generator, $g$, as simple as possible. We use $g_1$ of FC-ReLU-FC with hidden $D=32$. We use Euclidean distance for $d$ and set hyperparameter $\lambda=0.1$
% and $\alpha=0.1$ 
as a default setting. During training, the softmax temperatures $\tau_{n\neq N+1}$ are fixed to 1, and $\gamma=3$, i.e., $\tau_{N+1}=3$, for $\mathcal{L}_{CE}$. The $\tau$ in Gumbel softmax is cosine annealed from 2 to 0.5 in Eq.~\ref{eq:gumbel} following~\cite{gumbel}. We use $L=3$ dummies as default.

\begin{table}[t]
    \caption{\textbf{splitGSC:} 5-way \{1, 5\}-shot FSOSR results. The numbers are mean (std) over 5 trials (\%). (\textbf{bold}: best)}
    \vskip -0.12in
    \label{table:main_result}
    \centering
    \resizebox{\linewidth}{!}{
    \begin{tabular}{lccccc}
    \toprule
    && \multicolumn{2}{c}{1-shot} & \multicolumn{2}{c}{5-shot} \\
    \cmidrule{3-4} 
    \cmidrule{5-6}
    Model & Backbone & Accuracy & AUROC & Accuracy & AUROC \\
    \midrule
    \midrule
    ProtoNet & Conv4-64 & 43.2 (0.7) & 55.3 (0.5) & 67.6 (1.0) & 63.8 (0.6)\\
    % OpenMax & Conv4-64 & \\
    
    FEAT & Conv4-64 & 46.9 (1.2) & 61.1 (0.3) & 65.6 (1.3) & 65.5 (0.7) \\
    PEELER & Conv4-64 & 42.8 (1.5) & 57.5 (1.0) & 66.3 (1.8) & 66.7 (1.0)\\
    
    SnaTCHer-F & Conv4-64 & \textbf{47.3 (1.2)} & 51.1 (0.7) & 67.0 (1.6) & 53.1 (1.4) \\
    % ~~ distance metric only & Conv4-64 & 43.2 (0.7) & 53.5 (1.1) & 67.6 (1.0) & 56.5 (1.4)\\
    
    \midrule
    D-ProtoNet, L=3 & Conv4-64 & 45.3 (0.9) & \textbf{65.9 (0.3)} & \textbf{69.6 (0.8)} & \textbf{73.9 (0.7)} \\
    % D-ProtoNet + FEAT & Conv4-64 & 49.0 (0.5) & 68.5 (0.2) & 68.1 (0.7) & 75.8 (1.0) \\
    \midrule
    \midrule
    ProtoNet & ResNet-12 & 68.3 (1.0) & 60.7 (0.7) & 85.9 (0.7) & 68.4 (1.2)\\
    % OpenMax & ResNet-12 & \\
    
    FEAT & ResNet-12 & 68.8 (1.1) & 65.6 (0.7) & 84.0 (0.9) & 71.0 (0.7)\\
    PEELER & ResNet-12 & 65.4 (1.5) & 66.2 (0.9) & 82.4 (1.1) & 72.1 (1.7)\\
    
    SnaTCHer-F & ResNet-12 & \textbf{70.4 (1.0)}& 73.2 (1.0) & 84.7 (0.6) & 83.3 (0.8)\\
    % ~~ distance metric only & ResNet-12 & 68.3 (1.0) & 72.0 (0.6) & 85.9 (0.7) & 80.3 (1.0) \\
    \midrule
    D-ProtoNet, L=3 & ResNet-12 & 69.7 (1.0) & \textbf{78.8 (0.3)} & \textbf{86.9 (0.3)} & \textbf{86.7 (0.3)}\\
    % D-ProtoNet + FEAT & ResNet-12 & 70.0 (1.3) & 76.5 (0.8) & 84.3 (1.3) & 82.3 (2.3)\\
    \midrule
    \midrule
    ProtoNet & BCResNet-8 & \textbf{66.7 (0.7)} & 60.9 (0.7) & \textbf{83.1 (0.5)} & 68.8 (0.6)\\
    % FEAT & BCResNet-8 & 64.7 (1.3) & 63.6 (0.5) & 79.5 (1.0) & 68.5 (1.2)\\
    D-ProtoNet, L=3 & BCResNet-8 & 65.2 (1.4) & \textbf{75.7 (0.9)} & 81.9 (1.1) & \textbf{82.3 (1.5)}\\
    \bottomrule
    \end{tabular}
    }
    \vskip -0.25in
\end{table}

\noindent \textbf{Baselines.} We compare our method to other notable approaches: ProtoNet~\cite{protonet}, FEAT~\cite{FEAT}, PEELER~\cite{PEELER}, and SOTA FSOSR method, SnaTCHer~\cite{snatcher_cvpr21} based on their available official implementations. There are the following changes to fit the baselines better to our \textit{split}GSC carefully. We set the hidden dimension of 16 for the Transformers in FEAT and SnaTCHer-F. Also, we set dropout rates in the Transformers of 0.5 and 0 for Conv4-64 and 0.6 and 0.1 for ResNet-12 experiments.

\subsection{Results}

Table~\ref{table:main_result} shows overall results in \textit{split}GSC. `Acc' stands for FSL accuracy, and we use the threshold-free area under the receiver-operating characteristics (AUROC) as an OSR measure following previous FSOSR approaches~\cite{PEELER,snatcher_cvpr21}. By introducing dummy prototypes and additional losses to various backbones, our D-ProtoNet significantly improves vanilla ProtoNet in AUROC and shows better FSL accuracies. Other baselines also improve vanilla ProtoNet, but D-ProtoNets shows clear margins. SnaTHCer is the strong baseline, and they suggest directly using distance metric instead of softmax output to detect open-set, i.e., $\bm{x}_i$ is an open-set sample if $\max_{n\in {1, \cdots, N}}\{-d(f_\phi(\bm{x}_i), \bm{c}_n)\} < \delta$ while other approaches~\cite{PEELER, openmax, OSR_example2} usually use $\max_{n\in {1, \cdots, N}}p(y_i=n|\bm{x}_i)$. However, we observed that the unnormalized distance metric does not always hold for detecting open-set and shows poor AUROC with Conv4-64 in \textit{split}GSC and our training details.

\begin{table}[t]
    \caption{\textbf{\textit{mini}ImageNet:} 5-way \{1, 5\}-shot FSOSR results. The numbers are mean (std) over 5 trials (\%). *SnaTCHer results are quoted from the paper.}
    \vskip -0.11in
    \label{table:miniimagenet}
    \centering
    \resizebox{\linewidth}{!}{
    \begin{tabular}{lccccc}
    \toprule
    && \multicolumn{2}{c}{1-shot} & \multicolumn{2}{c}{5-shot} \\
    \cmidrule{3-4} 
    \cmidrule{5-6}
    Model & Backbone & Accuracy & AUROC & Accuracy & AUROC \\
    \midrule
    \midrule
    ProtoNet & ResNet-12 & 63.1 (0.4) & 55.3 (0.1) & 82.2 (0.2) & 61.0 (0.4) \\
    % OpenMax & ResNet-12 &\\
    
    FEAT & ResNet-12 & 67.1 (0.4) & 59.6 (0.5) & 82.3 (0.4) & 62.5 (0.2)\\
    PEELER & ResNet-12 & 62.1 (0.2) & 59.2 (0.3) & 81.8 (0.2) & 67.4 (0.3)\\
    SnaTCHer-F & ResNet-12 & 67.4 (0.4) & 69.4 (0.8) & \textbf{82.4 (0.1)} & 76.4 (0.7)\\
    ~~ *SnaTCHer-F~\cite{snatcher_cvpr21} & ResNet-12 & 67.0 & 68.3 & 82.0 & 77.4 \\
    ~~ *SnaTCHer-T~\cite{snatcher_cvpr21} & ResNet-12 & 66.6 & \textbf{70.2} & 81.8 & 76.7 \\
    ~~ *SnaTCHer-L~\cite{snatcher_cvpr21} & ResNet-12 & \textbf{67.6 }& 69.4 & \textbf{82.4} & 76.2\\
    % ~~ distance metric only & ResNet-12 & 63.1 (0.4) & 67.9 (0.4) &  82.2 (0.2)& 74.6 (0.7) \\
    
    \midrule
    D-ProtoNet, L=1 & ResNet-12 & 63.2 (0.3) & 69.7 (0.3) & 82.1 (0.1) & 76.6 (0.5)\\
    D-ProtoNet, L=3 & ResNet-12 & 63.4 (0.4) & \textbf{70.2 (0.4)} & 81.8 (0.2) & \textbf{77.8 (0.3)} \\
    \midrule
    \midrule
    D-ProtoNet, L=3 + FEAT & ResNet-12 & 65.1 (0.2) & 70.6 (0.6) & 81.9 (0.1) & 78.5 (0.9)\\
    \bottomrule
    
    \end{tabular}
    }
    \vskip -0.25in
\end{table}

\subsection{miniImageNet}

We further verify D-ProtoNets on widely used \textit{mini}-ImageNet~\cite{matchingnet_conv4_64_miniimagenet}. The dataset is the subset of ImageNet~\cite{ILSVRC15} and contains 100 classes with 600 84$\times$84 images for each. Following~\cite{ravi_split_iclr17}, we split the data into train, validation, and test sets with 64, 16, and 20 classes, respectively. We follow the implementations and training details of FEAT and SnaTCHer's official implementations for \textit{mini}ImageNet. The backbone is the ResNet-12 introduced by \cite{resnet12_mod}, whose output dimension is 640. Backbones are pretrained by adding a softmax layer to classify all 64 seen classes~\cite{ FEAT, snatcher_cvpr21}. Then, the model is trained for 200 epochs with 100 episodes for each. In each 5-way episode, we randomly choose 5 open-set classes, and 15 queries for all classes. We use softmax temperature $\tau=64$ for FEAT and SnaTCHer, following their best settings. We optimized $\tau$ of vanilla ProtoNet in $\{0.1, 1, 16, 32, 64, 128\}$ following \cite{FEAT} and found that $\tau=16$ works best. D-ProtoNets use $\tau_{n\neq N+1}=16$ and $\gamma=3$.

Table~\ref{table:miniimagenet} shows the comparison between D-ProtoNets and various baselines. Our dummy prototypes improve vanilla ProtoNets in detecting open-set samples while preserving FSL accuracies, and achieve SOTA level open-set detection. FEAT and SnaTCHer show better FSL accuracies than ours, especially for 1-shot settings. We interpret that their additional set-to-set transformation for support examples makes the extreme low shot regime more robust by using the relation between a few support examples. However, both methods do additional transformations, and SnaTCHer requires heavy computations due to the transformations for all the queries.

\bg{We also tried the transformation method FEAT to our D-ProtoNets to improve further. At inference time, we observed that using an unnormalized distance metric, $-d(f_\phi(\bm{x}_i), \bm{c}_{N+1})$, instead of softmax output~\cite{snatcher_cvpr21} increases the 1-shot and 5-shot AUROC from 68.5 to 70.6 and from 78.2 to 78.5, respectively, while using the D-ProtoNets+FEAT. The result implies that D-ProtoNets can be used with a complementary concept like FEAT.}

\subsection{Ablation Study}

\begin{table}[t]
    \caption{\textbf{Ablations.} splitGSC: 5-way \{1, 5\}-shot FSOSR results using ResNet-12. The numbers are mean (std) over 5 trials (\%). `Gum.' stands for Gumbel softmax.}
    \vskip -0.12in
    \label{table:ablation}
    \centering
    \resizebox{0.9\linewidth}{!}{
    \begin{tabular}{cccccccc}
    \toprule
    \multicolumn{4}{c}{Methods} & \multicolumn{2}{c}{1-shot} & \multicolumn{2}{c}{5-shot} \\
    \cmidrule{5-6} 
    \cmidrule{7-8}
    L & $\mathcal{L}^U_{CE}$ & $\gamma$ & Gum. & Accuracy & AUROC & Accuracy & AUROC \\
    \midrule
    \midrule
    - & & - & - & 68.3 (1.0) & 60.7 (0.7) & 85.9 (0.7) & 68.4 (1.2) \\
    3 & \checkmark & 1 & & \textbf{69.9 (1.1)} & 76.4 (1.1) & 86.4 (1.1) & 82.7 (1.2) \\
    3 & \checkmark & 3 & & 69.2 (1.4) & 78.7 (0.9) & 86.5 (0.7) & 86.2 (0.9) \\
    3 & \checkmark & 3 & \checkmark& 69.7 (1.0) & \textbf{78.8 (0.3)} & \textbf{86.9 (0.3)} & \textbf{86.7 (0.3)}\\
    \midrule
    1 & \checkmark & 3 & - & 70.0 (0.5) & 78.7 (0.1) & 86.7 (0.3) & 85.9 (0.4)\\
    2 &\checkmark & 3 & \checkmark & 69.1 (0.5) & 78.5 (0.4) & 86.4 (0.2) & 86.1 (0.6) \\
    3 & \checkmark& 3 & \checkmark&  69.7 (1.0) & \textbf{78.8 (0.3)} & \textbf{86.9 (0.3)} & \textbf{86.7 (0.3)}\\
    4 & \checkmark& 3 & \checkmark& \textbf{70.3 (0.7)} &\textbf{ 78.8 (0.6)} & \textbf{86.9 (0.7)} & 86.5 (1.2) \\
    5 & \checkmark& 3 & \checkmark& 69.2 (0.5) & 78.4 (0.7) & 86.4 (0.4) & \textbf{86.7 (0.5)} \\
    \midrule
    3 & \checkmark& 1 & \checkmark& 70.1 (2.0) & 76.8 (0.8) & 86.4 (1.1) & 82.2 (0.6) \\
    3 & \checkmark& 2 & \checkmark& 69.8 (1.0) & 78.3 (0.4) & 87.0 (0.7) & 85.9 (0.6) \\
    3 & \checkmark& 3 & \checkmark&  69.7 (1.0) & 78.8 (0.3) & \textbf{86.9 (0.3)} & \textbf{86.7 (0.3)}\\
    3 & \checkmark& 5 & \checkmark& 69.8 (1.1) & 78.7 (0.6) & 86.5 (0.6) & 86.4 (0.8)\\
    3 & \checkmark& 10 & \checkmark& \textbf{70.3 (1.0)} & \textbf{79.1 (0.5)} & 86.3 (0.6) & \textbf{86.7 (0.3)}\\
    \bottomrule
    \end{tabular}
    }
    \vskip -0.10in
\end{table}

\noindent \textbf{Ablations.} We do some ablation studies for D-ProtoNets, and Table~\ref{table:ablation} shows the effect of our loss terms, number of dummies, $\tau_{N+1}$, and Gumbel softmax. The top row implies the baseline, ProtoNet. By adding dummies, there are considerable improvements of more than 10 \% in AUROC. We could get further improvements through introducing Gumbel softmax and the $\tau_{N+1}=3$ while $\tau_{n\neq N+1}$ are fixed to 1. We also experiment with how the number of dummies, L, affects D-ProtoNets. D-ProtoNets shows better results as L increases from $1$ to $3$ and does not show further improvements with $L=4$ and $5$ (seem converged at $L=3$). We further experiment the effect of various $\gamma\in\{1, 2, 3, 5, 10\}$ for $\tau_{N+1}$. The AUROC increases as $\gamma$ increases from $1$ to $3$ and seems to converge near $\gamma=3$. 

\noindent \textbf{Erase undesirable instance discrepancy.} Recently, \cite{COSOC_neurips21} shows the importance of making few-shot models concentrate on foreground objects rather than backgrounds in images. Motivated by \cite{COSOC_neurips21}, we use an explicit normalization along the frequency-axis, named Relaxed instance Frequency-wise Normalization (RFN)~\cite{RFN, resnorm_dcase21_workshop}, to reduce undesirable instance discrepancy in audio features. The RFN module operates its input $\bm{x}$ and outputs $\lambda\cdot \text{LN}(\bm{x}) + (1-\lambda)\cdot \text{IFN}(\bm{x})$, where IFN is instance normalization~\cite{IN} along frequency-axis, and LN is layer normalization~\cite{LN} for relaxing the effect of IFN. Here we use the relaxation $\lambda=0.5$ and apply RFN at the input of the encoder $f_\phi$. We expect RFN to make the model concentrate on keywords more than other discrepancies, e.g., speaker ID. Table~\ref{table:RFN} shows that RFN consistently improves ProtoNet and D-ProtoNet with various backbones. The results indicate that erasing undesirable instance discrepancy is highly important in FSOS-KWS. \nocite{rfn_dcase_techreport}

\begin{table}[t]
    \caption{\textbf{Using RFN.} splitGSC: 5-way \{1, 5\}-shot FSOSR results. The numbers are mean (std) over 5 trials (\%).}
    \vskip -0.12in
    \label{table:RFN}
    \centering
    \resizebox{0.95\linewidth}{!}{
    \begin{tabular}{lccccc}
    \toprule
    && \multicolumn{2}{c}{1-shot} & \multicolumn{2}{c}{5-shot} \\
    \cmidrule{3-4} 
    \cmidrule{5-6}
    Model & Backbone & Accuracy & AUROC & Accuracy & AUROC \\
    \midrule
    \midrule
    ProtoNet & Conv4-64 & 43.2 (0.7) & 55.3 (0.5) & 67.6 (1.0) & 63.8 (0.6)\\
    ~~ + RFN & Conv4-64 & \textbf{46.3 (0.9)} & \textbf{57.3 (0.7)} & \textbf{69.9 (1.0)} & \textbf{65.3 (0.6)}\\
    \midrule
    D-ProtoNet, L=3 & Conv4-64 & 45.3 (0.9) & 65.9 (0.3) & 69.6 (0.8) & 73.9 (0.7) \\
    ~~ + RFN & Conv4-64 & \textbf{48.8 (0.6)} & \textbf{69.3 (0.5)} & \textbf{71.9 (0.7)} & \textbf{76.9 (0.4)}\\
    
    \midrule
    \midrule
    ProtoNet & ResNet-12 & 68.3 (1.0) & 60.7 (0.7) & 85.9 (0.7) & 68.4 (1.2)\\
    ~~ + RFN & ResNet-12 & \textbf{70.5 (1.0)} & \textbf{62.9 (0.7)} & \textbf{87.3 (0.8)} & \textbf{70.9 (1.4)}\\
    \midrule
    D-ProtoNet, L=3 & ResNet-12 & 69.7 (1.0) &     78.8 (0.3) & 86.9 (0.3) & 86.7 (0.3) \\
    ~~ + RFN & ResNet-12 & \textbf{72.6 (0.5) }& \textbf{80.3 (1.0)} & \textbf{88.3 (0.6)} & \textbf{87.8 (0.9)}\\
    \midrule
    \midrule
    ProtoNet & BCResNet-8 & 66.7 (0.7) & 60.9 (0.7) & 83.1 (0.5) & 68.8 (0.6)\\
    ~~ + RFN & BCResNet-8 & \textbf{71.1 (1.6)} & \textbf{65.2 (1.0)} & \textbf{86.3 (0.7)} & \textbf{74.2 (1.0)}\\
    \midrule
    D-ProtoNet, L=3 & BCResNet-8 & 65.2 (1.4) & 75.7 (0.9) & 81.9 (1.1) & 82.3 (1.5) \\
    ~~ + RFN & BCResNet-8 & \textbf{69.7 (0.5)} & \textbf{78.3 (0.3)} & \textbf{85.5 (0.3)} & \textbf{85.4 (0.6)} \\
    \bottomrule
    \end{tabular}
    }
    \vskip -0.25in
\end{table}

\section{Conclusion}

This work tackles few-shot open-set recognition in keyword spotting (FSOS-KWS) and suggests a new benchmark, \textit{split}GSC. To adapt to the varying open-set, we introduce episode-known dummies to Prototypical Networks (ProtoNets), named Dummy Prototypical Networks (D-ProtoNets). D-ProtoNets shows clear margins compared to recent baselines in \textit{split}GSC and achieves SOTA open-set detection in \textit{mini}ImageNet. We also suggest future research, erasing undesirable instance discrepancy, for FSOS-KWS.

\bibliographystyle{IEEEtran}

\bibliography{mybib}

% \begin{thebibliography}{9}
% \bibitem[1]{Davis80-COP}
%   S.\ B.\ Davis and P.\ Mermelstein,
%   ``Comparison of parametric representation for monosyllabic word recognition in continuously spoken sentences,''
%   \textit{IEEE Transactions on Acoustics, Speech and Signal Processing}, vol.~28, no.~4, pp.~357--366, 1980.
% \bibitem[2]{Rabiner89-ATO}
%   L.\ R.\ Rabiner,
%   ``A tutorial on hidden Markov models and selected applications in speech recognition,''
%   \textit{Proceedings of the IEEE}, vol.~77, no.~2, pp.~257-286, 1989.
% \bibitem[3]{Hastie09-TEO}
%   T.\ Hastie, R.\ Tibshirani, and J.\ Friedman,
%   \textit{The Elements of Statistical Learning -- Data Mining, Inference, and Prediction}.
%   New York: Springer, 2009.
% \bibitem[4]{YourName17-XXX}
%   F.\ Lastname1, F.\ Lastname2, and F.\ Lastname3,
%   ``Title of your INTERSPEECH 2022 publication,''
%   in \textit{Interspeech 2022 -- 23\textsuperscript{rd} Annual Conference of the International Speech Communication Association, September 18-22, Incheon, Korea, Proceedings, Proceedings}, 2022, pp.~100--104.
% \end{thebibliography}

\end{document}